\documentstyle[twocolumn,aps,prd,epsf,epsfig]{revtex}

\begin{document}
\twocolumn[\hsize\textwidth\columnwidth\hsize
\csname@twocolumnfalse%
\endcsname
\draft
\title{Nonexponential motional damping of impurity atoms 
in Bose-Einstein condensates}
\author{I.E.~Mazets$^{1,2} $ and G.~Kurizki$^1$}
\address{{\setlength{\baselineskip}{18pt}
$^1$\,Department of Chemical Physics, Weizmann Institute of Science, 
Rehovot 76100, Israel,\\
$^2$\,Ioffe Physico-Technical Institute, St.Petersburg 194021, Russia\\
}} 
\maketitle

\begin{abstract}
We demonstrate that the damping of the motion 
of an impurity atom injected at a 
supercritical velocity into a Bose-Einstein condensate can 
exhibit appreciable deviations from the exponential law on 
time scales of $10^{-5}$~s.  
\\     
\end{abstract}
\vskip1pc]

The decay of an unstable quantum system has long been known to 
deviate from exponential law for both 
very short and very long times \cite{std}~-- \cite{otk}. 
The short-time deviation from the exponential decay gives rise to either 
slowdown or speedup of the decay by frequently repeated measurements, 
known, respectively, as the quantum Zeno effect (QZE) \cite{std} and 
the anti-Zeno effect (AZE) \cite{QZAZ,AZKK}. A general unified 
theory of the QZE and AZE has been given in Ref.\cite{otk}. Among the 
proposed realizations of the QZE/AZE are photodetachement 
of negative ions \cite{rzh} and radiative decay of excited atoms in 
cavities \cite{AZKK}, photonic band gap structures \cite{pbg} 
or in the presense of a Bose-Einstein condensate (BEC) \cite{msh}. 
The first unambiguous 
observation of nonexponential decay in an unstable quantum system has 
been reported in  \cite{R0}, followed by the demonstration of 
the QZE and AZE \cite{R1}. These experiments 
have measured the escape of cold atoms 
from wells of an accelerating periodic potential induced by a 
standing light wave with varying frequency. 
Qualitatively,  escape from a trapping potential 
resembles nuclear alpha-decay  
\cite{Gamow} rather than decay via quanta emission into a bath, 
as in radiative decay \cite{Heitler} or in beta-decay \cite{Fermi}. 

The difficulty impeding the demonstration of nonexponential decay 
via quanta emission has been the short non-Markovian (memory) time of 
the relevant (electromagnetic or leptonic) 
bath, which is the time-scale of the effect \cite{otk}. 
Here we propose the emission of 
sound quanta (phonons) by atoms into a BEC as a candidate process 
for the observation of nonexponential relaxation, 
taking advantage of the rather long memory time of the condensate. The 
detailed understanding of such decoherence processes is important for the 
envisaged use of atomic BECs in metrology and interferometry \cite{mi}.  

Let us consider an 
``impurity'' atom moving in an atomic BEC. Atoms of another isotope or 
the same isotope but in a different internal (hyperfine) state can be 
viewed as impurities as long as their density is small enough not to 
modify considerably the BEC excitation spectrum. At  
``supercritical'' velocities, namely, above 
the speed of sound in the BEC, the impurity atom is decelerated 
due to phonon creation in the BEC. The rate of such a 
process according to the standard Fermi golden rule (i.e., assuming 
exponential decay of the amplitude of the initial state) has 
been calculated for both a uniform BEC \cite{TC} and a 
harmonically trapped BEC \cite{idz} and has been determined experimentally
\cite{ek}. We shall step beyond this approach and calculate more 
generally the time evolution of the impurity-atom motion in such a system. 

The state vector of the system can be written as (we set $\hbar =1$)  
\begin{equation} 
\left| \psi (t)\right \rangle =\alpha _{in}(t)\left| {\mathrm in}
\right \rangle +\int \frac {d^3{\bf q}}{(2\pi )^3}\, \beta _q(t) 
\left| {\bf q}\right \rangle , 
\label{statev}
\end{equation} 
where the initial state $\left| {\mathrm in}
\right \rangle $ corresponds to an impurity atom of mass $m_1$ 
moving at the velocity ${\bf V}_1$ in a BEC with no 
elementary excitation, and $\left| {\bf q}\right \rangle $ denotes the 
state where one elementary excitation of the BEC 
with the momentum {\bf q} is 
present (correspondingly, the impurity momentum is changed to 
$m_1{\bf V}_1-{\bf q}$). The initial conditions are, naturally,
\begin{equation} 
\alpha _{in}(0)=1,\quad \beta _q(0)=0.
\label{icond}
\end{equation} 
The set of equations describing motional damping of the 
impurity atom is 
\begin{eqnarray} 
i\dot \alpha _{in}&=&\left( \frac {m_1V_1^2}2+
\tilde{g}_{12}n\right) \alpha _{in}+ \nonumber \\ &&
\tilde{g}_{12}\sqrt{n} 
\int \frac {d^3{\bf q}}{(2\pi )^3}\, (u_q-v_q)\beta _q,  
\label{eqmo1}   \\
i\dot \beta _q&=&\left[ \frac {(m_1{\bf V}_1-{\bf q})^2}{2m_1}+
\epsilon (q) +\tilde{g}_{12}n\right] \beta _q+ \nonumber \\ && 
\tilde{g}_{12}\sqrt{n}(u_q-v_q)\alpha _{in}. 
\label{eqmo2}
\end{eqnarray} 
Here $n$ is the uniform BEC density, 
$\tilde{g}_{12}$ is the effective interspecies 
coupling constant, $u_q=\{ [{E_{HF}(q)}/{\epsilon (q)}+1]/2\} ^{1/2}$ 
and  $v_q=(u_q^2-1)^{1/2}$ are the Bogoliubov 
transformation coefficients, 
and $\epsilon (q)=[E_{HF}^2(q)-\mu^2] ^{1/2}$ 
is the energy of the elementary 
excitation with momentum {\bf q} \cite{seb}.  
Here we have introduced the Hartree-Fock excitation energy 
$E_{HF}=q^2/(2m_2)+
\mu $ and the chemical potential of the BEC $\mu =4\pi a_{22}n/m_2$, 
$a_{22}$ being the intraspecies {\em s}-wave scattering length for the 
condensed atoms of mass $m_2$. 
Correspondingly, the speed of sound in the BEC is 
$c_s=\sqrt{\mu /m_{2}}$. 

To proceed, we have to reckon with the 
renormalization of the interspecies coupling constant
\cite{ren}. The renormalized constant $\tilde{g}_{12}$ is expressed in 
terms of the bare coupling constant $g_{12}$  as 
\begin{equation} 
\tilde g_{12}=g_{12}\left[ 1+2mg_{12}(2\pi )^{-3}\int d^3{\bf q}
\, q^{-2} \right] ,  \label{ren12} 
\end{equation} 
with $g_{12}=2\pi a_{12}/m$, 
$a_{12}$ being the interspecies {\em s}-wave scattering length 
and $m=m_1m_2/(m_1+m_2)$ being the reduced mass. In the approximate 
(perturbative) solution of Eqs. (\ref{eqmo1}, \ref{eqmo2}) we shall 
keep the terms up to the second order in the {\em bare} 
constant $g_{12}$. 

The easiest way to solve Eqs. (\ref{eqmo1}, \ref{eqmo2}) 
is by the Laplace transformation (cf. Ref.\cite{pbg}). 
We adopt the interaction representation, wherein the 
probability amplitude of the initial state is 
\begin{equation}
\alpha (t)=\alpha _{in}(t)\exp [i(m_1V_1^2/2+g_{12}n)t] .
\label{ir}
\end{equation} 
The algebraic solution for the Laplace transform of $\alpha (t)$,  
$\overline{\alpha }(s)=\int _0^\infty dt\, \exp (-st)\alpha (t)$ 
has the form  
\begin{equation}
\overline{\alpha }(s)=\left[ s+\overline{\Omega }(s)\right] ^{-1}, 
\label{sll} 
\end{equation}
\begin{equation}
\overline{\Omega }(s) =g_{12}^2n\int \frac {d^3{\bf q}}{(2\pi )^3}\, 
\left[ \frac {(u_q-v_q)^2}{s+i\Delta (q)}+\frac {2im}{q^2}\right] ,
\label{ggen}
\end{equation}
$\Delta (q)=\epsilon (q)+q^2/(2m_1)-{\bf qV}_1$ being the energy 
mismatch between the the states $\left| {\mathrm in}\right \rangle $ 
and $\left| {\bf q} \right \rangle $. The second term in the square 
brackets in Eq. (\ref{ggen}) arises from the coupling-constant 
renormalization in Eq.(\ref{ren12}) 
and compensates for the ultraviolet divergence of the 
first term. This compensation is completely analogous to that of the 
electron mass 
renormalization in calculations of the radiative shift of an atomic 
optical transition \cite{be}. 

Equations (\ref{eqmo1}, \ref{eqmo2}) yield over  
a broad time interval, excluding very short times,  
exponential decay of  
$\alpha (t)\propto \exp [-(\gamma /2+i\omega _s)t]$, with the rate  
$\gamma $ and the frequency shift $\omega _s$  
\begin{equation}
\gamma =\lim _{s\rightarrow 0} 2\, {\mathrm Re}\,
\overline{\Omega }(s), 
\quad 
\omega _s=\lim _{s\rightarrow 0} {\mathrm Im}\,
\overline{\Omega }(s).
\label{lpgo} 
\end{equation} 
The relaxation rate $\gamma $ can be calculated within the 
exact Bogoliubov theory using Fermi's golden rule \cite{TC,ek}.

It is possible to obtain a nonexponential  
analytical solution for $\alpha (t)$ in  
the Hartree-Fock (HF) approximation \cite{HF} (hereafter we label all 
the quantities in this approximation by HF): 
$\epsilon _{HF}(q)= E_{HF}(q)$, $u_{q\, HF}= 1$, $v_{q\, HF}= 0$. 
Then the integral in Eq.(\ref{ggen}) can be evaluated analytically, 
yielding in Eq. (\ref{lpgo})
the exponential decay rate  
$\gamma _{HF}=\gamma _0[1- c_{HF}^2/V_1^2]^{1/2}$ 
if $V_1>c_{HF}\equiv \sqrt{2\mu /m_2}$ and zero otherwise. Here  
$\gamma _0=4\pi a_{12}^2nV_1$         
is the collison rate calculated 
for the impurity atom using the Fermi golden rule in the limit 
$a_{22}\rightarrow 0$ (an ideal BEC).  Finally, in the case  
$V_1>c_{HF}$ we obtain  
\begin{equation}
\overline{\alpha }_{HF}(s)=\left[ \gamma _{HF}
\left( \frac s{\gamma _{HF}}+\frac 12 
\sqrt{1+\frac {is}{\gamma _{HF}K}} \right) \right] ^{-1}, 
\label{gal1} 
\end{equation}
where $K=(mV_1^2/2-\mu )/\gamma _{HF}$ is the energy of collision 
of the impurity atom with a condensate atom in 
their center-of-mass frame 
(including the correction due to the BEC chemical potential $\mu $) 
scaled to $\gamma _{HF}$. By inverting such a Laplace transform  
\cite{abr} we get 
\begin{equation}
\alpha _{HF}(t)=\frac {e^{i\Lambda t}}{\xi _1-\xi _2}\left[ 
\xi _1\varphi (\xi _1\sqrt{t}) -\xi _2\varphi (\xi _2\sqrt{t})
\right] ,
\label{ahf}
\end{equation} 
where $\varphi (z)=\exp (z^2){\mathrm erfc}(-z)$, 
${\mathrm erfc}(z)$ being the complementary error function, and 
$\xi _{1,2}=-\Xi /2 \pm \sqrt{\Xi ^2/4-i\Lambda }$ are the roots 
of the quadratic equation $\xi ^2+\Xi \xi +i\Lambda =0$.  
The coefficients of the latter equation are 
$\Xi =\gamma _0\sqrt{i/(2mV_1^2)}$ and $\Lambda =mV_1^2/2-\mu $. 

\begin{figure}
\centerline{\psfig{file=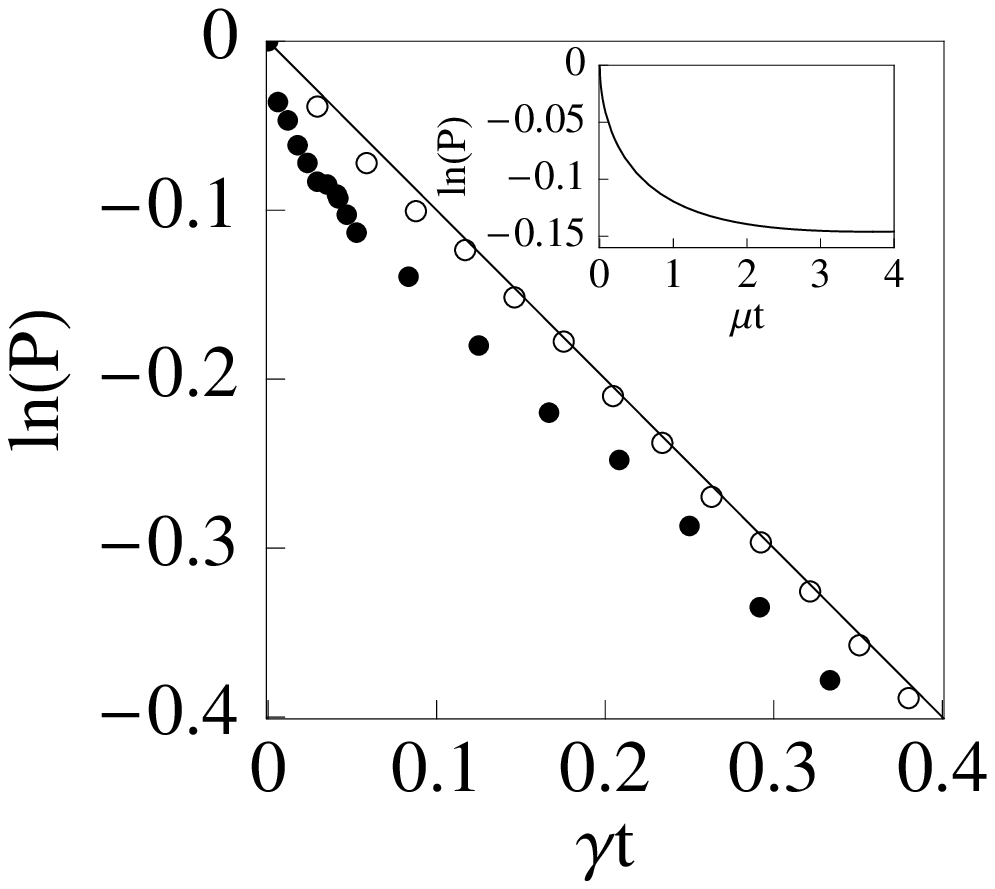,width=7.8cm}}
\vspace*{3mm}
\begin{caption} 
{Numerically calculated logarithm of survival probability 
$P(t)$ of 
the initial state of impurity atoms in a $^{87}$Rb BEC  
plotted versus  dimensionless time $\gamma t$ for 
$n=10^{14}$~cm$^{-3}$ ($c_s=0.2$~cm/s); $m_1=m_2$, $a_{12}=3a_{22}$. 
Filled circles: $V_1=3c_s$, $\gamma =1.1\cdot 10^3$~s$^{-1}$. 
Open circles: $V_1=7c_s$, $\gamma =4.4\cdot 10^3$~s$^{-1}$. 
Solid line: exponential law $\exp (-\gamma t)$.  Inset: 
typical behavior of logarithm of $P(t)$ for the case of subcritical 
$V_1$ (practically any  $V_1\le 0.7 c_s$). Note the different 
scaling (by $\mu $) of the horizontal axis of the inset plot.}
\end{caption}
\end{figure}

Although our numerical results (Fig.~1) show that 
the HF approximated solution Eq. (\ref{ahf})  
is rather crude, it nonetheless  provides a qualitative guidance to the 
physical behavior. Simple scaling considerations   
lead us to the conclusion that $\alpha _{HF}$ depends on two 
parameters: the dimensionless time variable $\gamma _{HF}t$ and 
$K$ [Eq. (\ref{gal1})].  
Equation (\ref{ahf}) predicts that at $t\rightarrow 0$ the 
decay is more rapid than exponential, so the survival 
probability $P(t)=|\alpha (t)|^2$ behaves in the HF approximation 
as  $P_{HF}(t)  \approx 1-4\, {\mathrm Re} \, 
(\xi _1+\xi _2)\sqrt{ t/\pi }$, becoming exponential at larger times,  
$P_{HF}(t)\approx \exp (-\gamma _{HF}t)$. 
The deviation from  exponential decay is appreciable 
only for $K\, ^<_\sim \, 1$. There are 
two ways to attain $K< 1$. One is to take a small difference 
between the impurity atom velocity $V_1$ and the critical velocity, but  
this would reduce the damping rate,  
which may be experimentally inconvenient. 
A much better way is to strive for a  large 
interspecies scattering length $a_{12}$, as discussed below. 

The results of our numerical calculations based on  
Eqs.~(\ref{sll},\, \ref{ggen}), which use the {\em exact} expressions 
for $\epsilon (q)$, $u_q$ and $v_q$ instead of the HF approximation, 
clearly reveal a deviation from exponential decay for small times.  
Under such conditions, frequent measurements would 
accelerate the decay, causing the anti-Zeno effect (AZE). Alternatively, 
one may accelerate the decay by 
periodically modulating the coupling of the initial state 
to the continuum \cite{otk}, instead of repeated projective 
measurements. This can be done by changing the impurity velocity using a 
sequence of Bragg or Raman laser pulses \cite{Bragg}. 

The inset to Fig.~1 shows that  if $V_1<c_s$, the survival probability 
first decreases  and then approaches the constant value of 
about 0.85, almost independent on $V_1$. This behavior reveals the 
physical reason for the short-time nonexponential decay: 
the initial conditions 
Eq.(\ref{icond}) imply that, initially, the impurity atom is 
surrounded by {\em no virtual phonons}, while in the steady state,  
the impurity atom must be surrounded by a cloud of virtual phonons 
(cf. the polaronic effect for electrons in a crystal \cite{isi}). 
Thus the nonexponential stage of the decay is  associated with the 
formation of such a phonon cloud. 

The faster the impurity atom moves, the weaker its 
coupling to the phonon cloud is. Therefore  
the decrease of $\omega _s$ corresponds to the  
vanishing of nonexponential decay effects as $V_1$ increases. This 
behavior is displayed in Fig.~2 by the numerically calculated 
[from Eq. (\ref{lpgo})]  
decay rate $\gamma $ and frequency shift $\omega _s$. 

Our numerical studies of Eqs. (\ref{eqmo1}, \ref{eqmo2}) always 
yield {\em decay acceleration} at short times, typically on 
the scale of $10^{-5}$~s. But  
should one not expect, from general considerations \cite{std}, 
$P(t)=1-{\mathrm const}\cdot t^2$ at $t\rightarrow 0$, 
in accordance with the QZE? To answer this question, 
we should apply the general theory developed in Ref.~ 
\cite{otk}, whereby the short-time behavior is determined by the 
spectrum (i.e., the dependence on the emitted quantum energy 
$\epsilon $) of the reservoir response $G(\epsilon )$. This 
spectrum is given by the  interaction matrix element 
squared multiplied by the density of 
the reservoir  states. In our case, we find that 
\begin{equation} 
G(\epsilon )=\left[ \frac {2\pi a_{12}(u_q-v_q)}m\right] ^2 
\frac {nq^2}{2\pi ^2}\frac {dq}{d\epsilon } 
\label{gna} 
\end{equation} 
monotonously increases with the emitted phonon energy $\epsilon $. 
According to Ref.~\cite{otk}, if the energy uncertainty $\sim t^{-1}$ 
associated with the finite observation time 
$t$ covers the energy range 
where $G(\epsilon )$ increases, then decay acceleration (AZE) takes 
place. However, our approach [Eqs.(\ref{eqmo1},\, \ref{eqmo2})] 
leading to Eq.(\ref{gna}) is valid only for small transferred 
momenta. If $q\,^>_\sim \, r_0^{-1}$, where $r_0\sim 10^{-7}$~cm 
is the characteristic radius of the interatomic potential, we 
cannot consider $a_{12}$ as constant any more. Instead, the 
interaction matrix element decreases with $q$ in this 
range. In the inset of Fig.~2 we 
schematically display the spectrum of $G(\epsilon )$, including 
its decreasing part, whose detailed calculation is beyond the scope of  
this Letter. This spectrum implies that at very short times 
($^<_\sim \, 10^{-9}$~s), the energy  
uncertainty broadening covers the whole profile of $G(\epsilon )$,  
thereby giving rise to the QZE. 

\begin{figure}
\centerline{\psfig{file=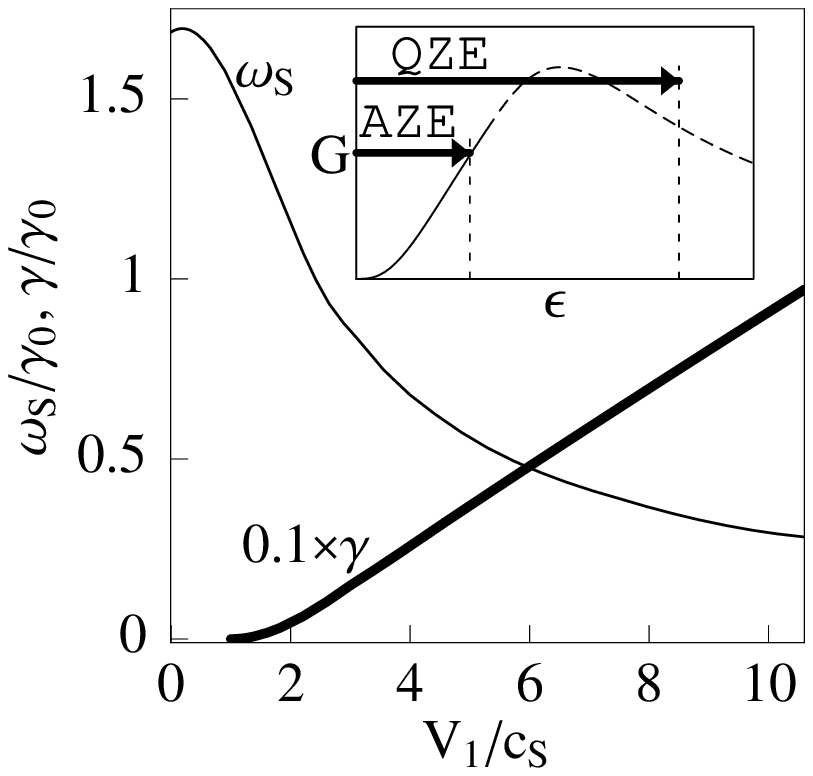,width=7.8cm}}
\vspace*{3mm}
\begin{caption} 
{Numerically calculated frequency shift $\omega _s$ and the 
exponential decay rate $\gamma $ (scaled to $\gamma _0$) 
versus the impurity 
dimensionless velocity $V_1/c_s$. Both  
$\omega _s/\gamma _0$ and $\gamma /\gamma _0$ 
display a universal behaviour, 
independent of the BEC density and the atomic species. Inset: 
schematic representation of the spectrum $G(\epsilon )$; solid 
line: the increasing part given by Eq.(\ref{gna}); dashed line: 
the remaining part,  which decreases 
at $\epsilon \, ^>_\sim \, 1/(mr_0^2)$. 
The arrows indicate ranges of the energy uncertainty 
corresponding to AZE and QZE.  }
\end{caption}
\end{figure}

The parameters used in Fig.~1 for a BEC of $^{87}$Rb may correspond to  
impurity atoms of the same isotope but in a 
different hyperfine state, obtained by a short Raman pulse. This 
method of impurity atom admixing conforms to the initial 
conditions of Eq.(\ref{icond}). However, to reach appreciable 
nonexponentiality by this method, one has 
to enhance interspecies scattering either by means of 
interspecies Feshbach resonance or  via 
laser-induced dipole-dipole interactions 
\cite{art}. Another possibility is the  use of 
two-isotope mixture, for example, 
a BEC of $^{87}$Rb atoms with admixed fermionic $^{40}$K  
atoms. Such a choice is of particular interest, since the 
large interspecies scattering length in this mixture \cite{KRb}, 
as well as the mass ratio between these two elements seem very 
promising for experimental search of nonexponential decay effects. 
We note that the condition on the impurity velocity in this case is 
opposite to that of the aforementioned 
case of impurities {\em generated} 
from a BEC by a Raman pulse. Indeed, the 
$^{40}$K atoms at rest co-exist with 
the $^{87}$Rb BEC for a time long 
enough to form virtual phonon clouds around 
them. If then the $^{40}$K atoms suddenly acquire, by 
the action of a Bragg 
or Raman pulse, the velocity $V_1\gg c_s$, the initial conditions, 
instead of Eq.(\ref{icond}), should assume the pre-existence of the 
phonon cloud (prior to the pulse), 
which vanishes when the impurities attain high (supercritical) 
velocities. Thus we expect that in the 
case of a two-element ultracold mixture, nonexponential decay 
features are most pronounced for $V_1\gg c_s$.   

To conclude, we have outlined the possibility of observing  
deviations from exponential decay for unstable momentum states  
of impurity atoms moving in a BEC with 
a supercritical velocity on time scales of $10^{-5}$~s. 
The effects of finite temperature and bosonic enhancement in 
this process  will be a subject of a separate work. 

We thank Dr. A.G.~Kofman for helpful discussions. 
The support of the German-Israeli Foundation, the EC 
(the QUACS RTN), ISF and Minerva is acknowledged. I.E.M. 
also thanks the RFBR (projects 02--02--17686, 03--02--17522).

\end{document}